\documentclass[fleqn,10pt]{wlscirep}
\usepackage{graphicx}
\usepackage{epstopdf}
\usepackage{color,soul}
\usepackage{hyperref}
\usepackage{lineno}
\usepackage{amsmath}
\usepackage{multirow}

\title{Galvanomagnetic properties of the putative type-II Dirac semimetal PtTe$_2$}

\author[1]{Orest Pavlosiuk}
\author[2,*]{Dariusz Kaczorowski}
\affil[1]{Institute of Low Temperature and Structure Research, Polish Academy of Sciences, P. O. Box 1410, 50-950 Wroc{\l}aw, Poland}
\affil[2]{Institute of Molecular Physics, Polish Academy of Sciences, Mariana Smoluchowskiego 17, 60-179 Pozna\'{n}, Poland}

\affil[*]{d.kaczorowski@int.pan.wroc.pl}

\begin{abstract}
Platinum ditelluride has recently been characterized, based on angle-resolved photoemission spectroscopy data and electronic band structure calculations, as a possible representative of type-II Dirac semimetals. Here, we report on the magnetotransport behavior (electrical resistivity, Hall effect) in this compound, investigated on high-quality single-crystalline specimens. The magnetoresistance (MR) of PtTe$_2$ is large (over $3000\%$ at $T=1.8$\,K in $B=9$\,T) and unsaturated in strong fields in the entire temperature range studied. The MR isotherms obey a Kohler's type scaling with the exponent $m$ = 1.69, different from the case of ideal electron-hole compensation. In applied magnetic fields, the resistivity shows a low-temperature plateau, characteristic of topological semimetals. In strong fields, well-resolved Shubnikov -- de Haas (SdH) oscillations with two principle frequencies were found, and their analysis yielded charge mobilities of the order of $10^3\,\rm{cm^2V^{-1}s^{-1}}$ and rather small effective masses of charge carriers, $0.11m_e$ and $0.21m_e$. However, the extracted Berry phases point to trivial character of the electronic bands involved in the SdH oscillations. The Hall effect data corroborated a multi-band character of the electrical conductivity in PtTe$_2$, with moderate charge compensation.
\end{abstract}
\begin{document}

\flushbottom
\maketitle
\thispagestyle{empty}

\section*{Introduction}

Topological semimetals (TSs) form an outstanding group of materials characterized by perfect linear dispersion of some bulk electronic states.\cite{Hasan2017,Armitage2017a} In accordance with presence or absence of Lorentz invariance, one discriminates type-I and type-II systems.\cite{Yan2017a} In the latter class of TSs, Dirac cone is strongly tilted with respect to Fermi level.\cite{Soluyanov2015a} Both type-I and type-II TSs can be composed from Dirac or Weyl fermions, depending on what kind of symmetries is preserved.\cite{Yang2014a} Nontrivial electronic structures of TSs give rise to unusual electronic transport properties, commonly considered being highly prospective for various electronic devices of new kind. A hallmark feature of TSs is chiral magnetic anomaly (CMA), which manifests itself as a negative magnetoresistance (MR) observed when electric and magnetic fields are collinear.\cite{Nielsen1983} Actually, negative MR was found in many TSs, more often in type-I materials \cite{Li2015c,Li2016,Hirschberger2016a,Niemann2016a,Xiong2015} but also in a few type-II systems.\cite{Lv2017} Another smoking-gun signature of TSs is the existence of topological surface states, which have a form of Fermi-arcs.\cite{Hasan2017,Armitage2017a} Their presence was confirmed experimentally in a large number of TSs, among them Cd$_3$As$_2$,\cite{Wang2016m} MoTe$_2$,\cite{Deng2016} WTe$_2$\cite{Li2017b} and TaAs.\cite{Lv2015}

Type-II topological semimetallic states have been revealed in several transition metal dichalcogenides and in MA$_3$ (M = V, Nb, Ta; A = Al, Ga, In) icosagenides.\cite{Chang2017a,Yan2017a,Zhang2017e,Deng2016,Soluyanov2015a,Autes2016,Huang2016b}
Within the former group of compounds, MoTe$_2$  and WTe$_2$ have been classified as type-II Weyl semimetals,\cite{Deng2016,Bruno2016} whereas platinum and palladium dichalcogenides have been established via angle-resolved photoemission spectroscopy (ARPES) experiments to represent the family of type-II Dirac semimetals.\cite{Yan2017a,Zhang2017e,Noh2017} The TSs nature of PdTe$_2$ is reflected in its peculiar magnetotransport behavior.\cite{Wang2016i,Fei2017} In turn, no comprehensive study on the electronic transport properties of the Pt-bearing counterpart has been reported in the literature up to date. In this work, we investigated the galvanomagnetic properties of PtTe$_2$ with the main aim to discern features appearing due to the alleged nontrivial topology of its electronic band structure.

\section*{Results and discussion}

\subsection*{Electrical resistivity, magnetic field-induced plateau and magnetoresistance}

\begin{figure}[h]
	\centering
	\includegraphics[width=17cm]{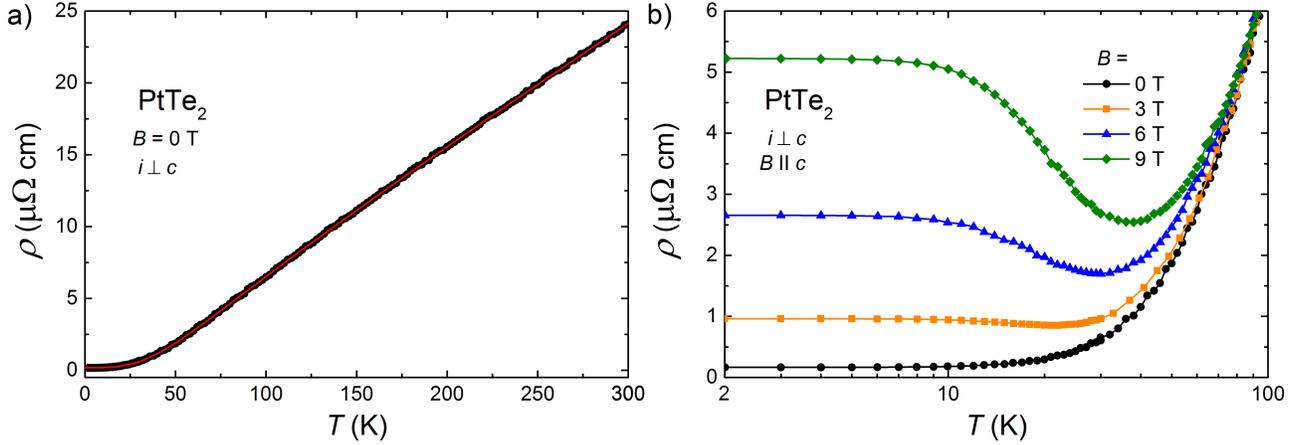}
	\caption{\textbf{Electrical resistivity of PtTe$_2$.} (a) Temperature dependence of the electrical resistivity measured in zero magnetic field with electric current $i$ confined within the $a-b$ plane of the crystallographic unit cell. Solid red curve represents the result of fitting the BG law to the experimental data. (b) Low-temperature resistivity ($i \perp c$) measured as a function of temperature in transverse magnetic field ($B \parallel c$).
	\label{rho(T)}}
\end{figure}

Figure~\ref{rho(T)}a shows the results of electrical resistivity, $\rho$, measurements performed on single-crystalline sample of PtTe$_2$, as a function of temperature, $T$, with electric current, $i$, flowing within the hexagonal $a-b$ plane. The overall behavior of $\rho(T)$ indicates a metallic character of the compound. The resistivity decreases from the value of $24.09\,\mu\Omega\,\rm{cm}$ at $T=300$\,K to $0.17\,\mu\Omega\,\rm{cm}$ at $T=2$\,K, yielding the residual resistivity ratio RRR = $\rho(300\,\rm{K})/\rho(2\,\rm{K})$ equal to 142. The magnitudes of both $\rho(2\,\rm{K})$ and RRR indicate high crystallinity of the specimen measured (RRR is approximately 5 times larger than that reported for PtTe$_2$ in Ref.\cite{Zhang2017e}). As displayed in Fig.~\ref{rho(T)}a (note the red solid line), in the whole temperature range covered, $\rho(T)$ can be very well approximated  with the Bloch-Gr{\"u}neisen (BG) law:
\begin{equation}
	\rho(T)=\rho_0+A\Bigg(\frac{T}{\varTheta_{\rm D}}\Bigg)^k\int_{0}^{\frac{\varTheta_{\rm D}}{T}} \frac{x^k}{(e^x-1)(1-e^{-x})} dx,
	\label{BG_eq}
\end{equation}
where $\rho_0$ is the residual resistivity, accounting for scattering conduction electrons on crystal imperfections, while the second term represents electron-phonon scattering ($\varTheta_{\rm D}$ stands for the Debye temperature). The least-squares fitting yielded the parameters: $\rho_0=0.17\,\mu\Omega\,\rm{cm}$, $\varTheta_{\rm D}=253$\,K, $A=11.6\,\mu\Omega\,\rm{cm}$ and $k=3.23$. The value of $k$ is smaller than $k=5$, expected for simple metals, yet similar to $k$ exponents determined for several monopnictides.\cite{Sun2016a,Pavlosiuk2017,Pavlosiuk2017b}

The temperature dependencies of the electrical resistivity of PtTe$_2$ measured in transverse magnetic field ($i \perp c$ axis and $B \parallel c$ axis) are gathered in Fig.~\ref{rho(T)}b. In non-zero $B$, $\rho(T)$ is a non-monotonous function of temperature, showing an upturn below a certain slightly field-dependent temperature $T*$, and then forming a plateau at the lowest temperatures. With increasing magnetic field, the magnitude of the resistivity in the turn-on and plateau regions distinctly increases. Similar behavior was considered as a fingerprint of the presence of nontrivial topology in the electronic structure of several TSs.\cite{Shekhar2015b,Li2016f,Singha2016a,Hosen2017} Another possible explanation of this type of magnetic field governed changes in $\rho(T)$ is a metal-insulator transition.\cite{Zhao2015,Li2016f,Du2005} However, as discussed first for WTe$_2$,\cite{Wang2015d} and afterwards, e.g., for rare-earth monopnictides,\cite{Zeng2016,Niu2016,Pavlosiuk2016f,Pavlosiuk2017,Pavlosiuk2017b,Kumar2016,Sun2016a,Ghimire2016,Han2017a,Xu2017f} the magnetic field induced upturn in $\rho(T)$ may appear also in trivial semimetals, which are close to perfect charge carriers compensation. We presume that the latter mechanism is also fully appropriate for the electrical transport behavior in PtTe$_2$.

\begin{figure}[h]
	\centering
	\includegraphics[width=17cm]{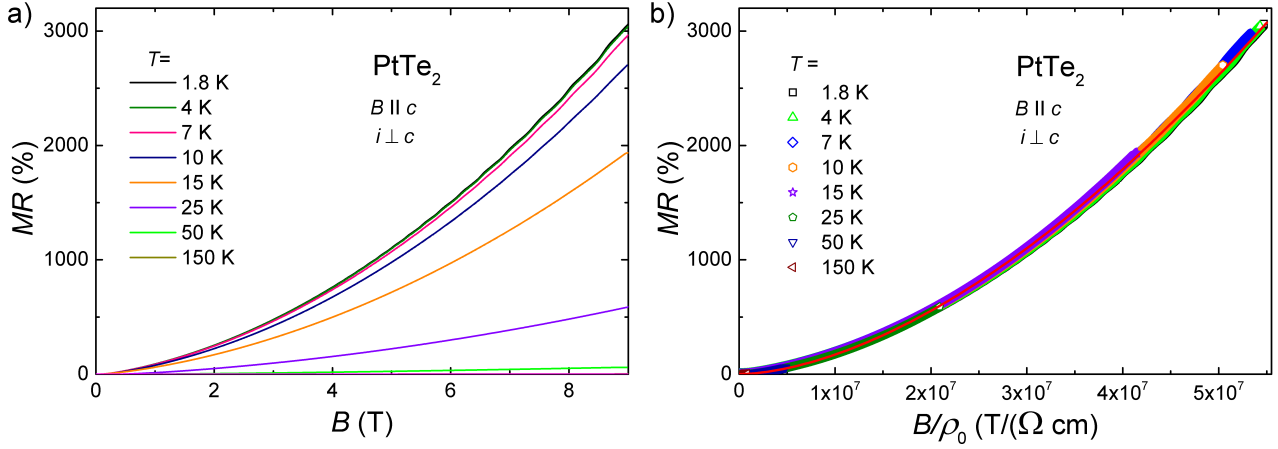}
	\caption{\textbf{Magnetotransport properties of PtTe$_2$. } (a) Magnetoresistance measured with $i \perp c$ and $B \parallel c$ at several constant temperatures. (b) Kohler's plot of the magnetoresistance data. Red solid line is the result of fitting the Kohler's equation to the experimental data.
	\label{mr(B)}}
\end{figure}

In order to examine the actual nature of the galvanomagnetic behavior observed for PtTe$_2$, transverse magnetoresistance, MR = $[\rho(B)-\rho(B=0)]/\rho(B=0)$, measurements were performed at several constant temperatures in the configuration $i \perp c$ axis and $B \parallel c$ axis. As can be inferred from Fig.~\ref{mr(B)}a, in $B=9$\,T, MR taken at $T=1.8$\,K achieves a giant value of $3060\%$, which is an order of magnitude larger than MR determined  in the same conditions for the related system PdTe$_2$,\cite{Wang2016i} and almost equal to that reported for the type-II Weyl semimetal MoTe$_2$.\cite{Chen2016a} With increasing temperature, MR measured in $B=9$\,T does not change significantly up to about 10 K (i.e., in the plateau region of $\rho(T)$), and then decreases rapidly. However, even at $T$ = 150 K, MR remains exceptionally large exceeding $500\%$ in 9~T. At each of the temperatures studied, MR shows no tendency towards saturation in strong fields. MR behavior similar to that of PtTe$_2$ was established before for several TSs.\cite{Wang2016i,Chen2016a,Shekhar2015b,Singha2016a,Kumar2017c,Wang2016l}
However, unsaturated MR can be also attributed to perfect or almost perfect carrier compensation in a semimetallic material.\cite{Ziman1972}

Remarkably, as demonstrated in Fig.~\ref{mr(B)}b, the MR isotherms of PtTe$_2$ obey the Kohler's rule in the entire temperature range studied. This finding rules out the scenario of metal-insulator transition as a possible mechanism of the magnetic field driven changes in the electrical transport of PtTe$_2$. The MR data collapse onto a single curve, which can be approximated by the expression: MR $\propto(B/\rho_0)^m$, with the exponent $m=1.69$ (note the red solid line in Fig.~\ref{mr(B)}b). While $m=2$ is expected for materials with perfect electron-hole balance, and the obtained value is also smaller than $m=1.92$ reported for WTe$_2$,\cite{Wang2015d} it is similar to those found for some monopnictides, which were reported as trivial semimetals fairly close to charge compensation.\cite{Pavlosiuk2016f,Pavlosiuk2017,Han2017a}

\subsection*{Quantum oscillations}

\begin{figure}[h]
	\centering
	\includegraphics[width=17cm]{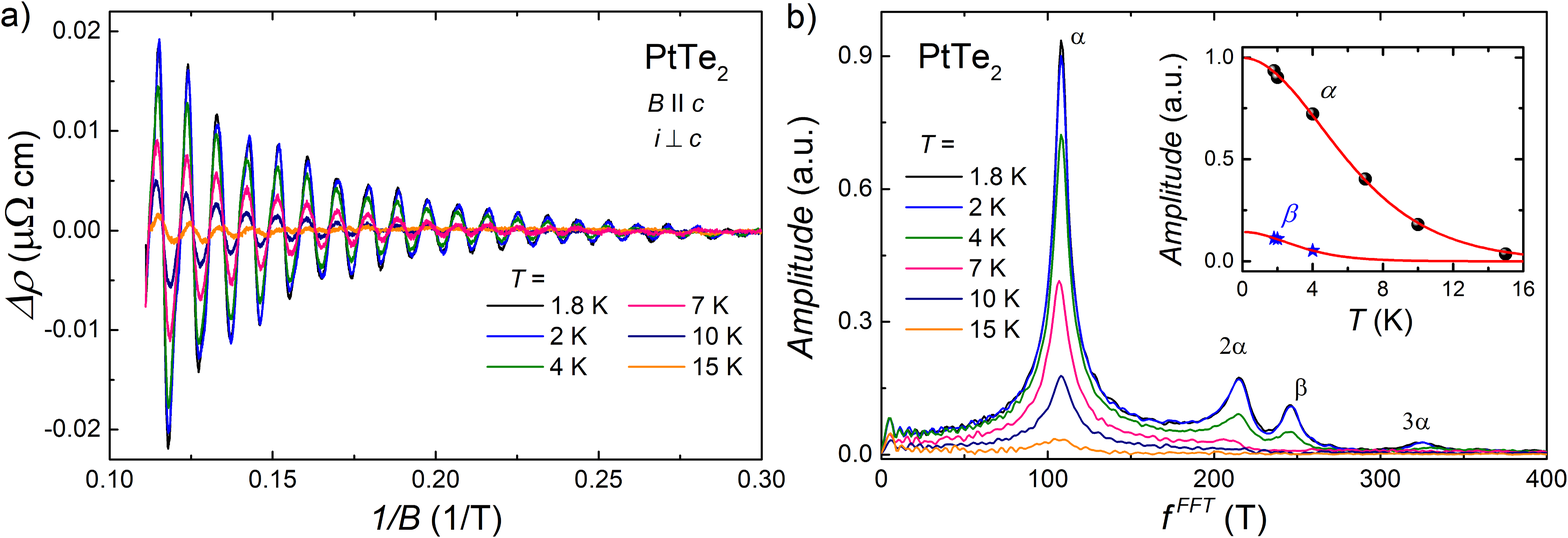}
	\caption{\textbf{Shubnikov-de Haas effect in PtTe$_2$.} (a) Oscillating part of the electrical resistivity measured at several different temperatures with $i \perp c$ axis and $B \parallel c$ axis, plotted as a function of inverse magnetic field. (b) Fast Fourier transform spectrum obtained from the analysis of the data presented in panel (a). Inset shows the temperature dependencies of the amplitudes of two principal frequencies in the FFT spectra. Solid lines represent the fits of Eq.~\ref{LK_eq} to the experimental data.
		\label{SdH_analysis}}
\end{figure}

In order to characterize the Fermi surface in PtTe$_2$, we investigated quantum oscillations in $\rho(B)$ (Shubnikov -- de Haas effect) at a few different temperatures. Fig.~\ref{SdH_analysis}a shows the oscillatory part of the electrical resistivity, $\Delta\rho$, obtained by subtraction of the second-order polynomial from the experimental data, plotted as a function of reciprocal magnetic field, $1/B$. As can be inferred from this figure, the SdH oscillations remain discernible at temperatures up to at least $15\,$K, however, their amplitudes systematically decrease with increasing temperature. Fast Fourier transform (FFT) analysis, the results of which are presented in Fig.~\ref{SdH_analysis}b, disclose four features at the oscillations frequencies $f_i^{\rm{FFT}}$ ($i$ represents the Fermi surface pocket label). The most prominent peak occurs at $f_{\alpha}^{\rm{FFT}}=108$\,T, and the corresponding Fermi surface pocket will be labeled thereinafter as $\alpha$. The next feature occurs at $f_{2\alpha}^{\rm{FFT}}=215$\,T and it is the second harmonic frequency of $f_{\alpha}^{\rm{FFT}}$. Then, the peak with its maximum at $f_{\beta}^{\rm{FFT}}=246$\,T can be attributed to another Fermi surface pocket, labeled $\beta$ in what follows. Eventually, the very weak maximum centered at $f_{3\alpha}^{\rm{FFT}}=325$\,T likely arises as the third harmonics of $f_{\alpha}^{\rm{FFT}}$. It is worthwhile noting that the FFT spectrum of PtTe$_2$ is very similar to that reported for PdTe$_2$ in Ref.\cite{Wang2016i} but differs from the FFT data shown for the same compound in Ref.\cite{Fei2017} Using the Onsager relation $f_i^{\rm{FFT}} = hS/e$, where $S$ stands for the area of Fermi surface cross-section, one finds for the two pockets in PtTe$_2$ the values $S_{\alpha}=1.03\times10^{-2}{\rm\AA^{-2}}$ and $S_{\beta}=2.34\times10^{-2}{\rm\AA^{-2}}$.
Assuming circular cross-sections, the corresponding Fermi wave vectors are $k_{\rm{F},\alpha}=5.73\times10^{-2}{\rm\AA^{-1}}$ and $k_{\rm{F},\beta}=8.63\times10^{-2}{\rm\AA^{-1}}$. Then, if we assume that both Fermi surface pockets are spherical, which is poor approximation, the carrier densities in these two pockets would be equal to $n_{\alpha}=6.35\times10^{18}$\,cm$^{-3}$ and $n_{\beta}=2.17\times10^{19}$\,cm$^{-3}$.

The electronic structure calculated for PtTe$_2$\cite{Yan2017a,Zhang2017e} comprises three bands crossing the Fermi level, one hole-like band, located at the center of the Brillouin zone, and two electron-like bands. The fact that we observed experimentally only two principle frequencies is probably due to very small size of the third Fermi surface pocket or might arise owing to somewhat lower position of the Fermi level in the single crystal studied. From the comparison of the values of $k_{\rm{F},i}$ obtained from the FFT analysis and sizes of the calculated electronic bands,\cite{Yan2017a} one may presume that the $\alpha$ Fermi surface pocket corresponds to one of the electron-like bands and the $\beta$ Fermi surface pocket represents the hole-like band.

The inset to Fig.~\ref{SdH_analysis}b displays the temperature variations of the FFT amplitudes, $R_i(T)$,  corresponding to the $\alpha$ and $\beta$ Fermi surface pockets in PtTe$_2$. The gradual damping of both oscillations with rising temperature can be described by formula:\cite{Shoenberg1984}

\begin{equation}
	R_i(T)\propto(\lambda m^*_iT/B_{\rm{eff}})/\sinh(\lambda m^*_iT/B_{\rm{eff}}),
	\label{LK_eq}
\end{equation}
where $m^*_i$ is the effective cyclotron mass of charge carriers, $B_{\rm{eff}}= 4.5$\,T was calculated as $B_{\rm{eff}}=2/(1/B_1+1/B_2)$ ($B_1=3$\,T and $B_2=9$\,T are the borders of the magnetic field window in which the FFT analysis was performed), and $\lambda=14.7\;$T/K was obtained from the relationship $\lambda=2\pi^2k_{\rm B}m_{\rm e}/e\hbar$ ($m_{\rm e}$ stands for the free electron mass, and $k_{\rm B}$ is the Boltzmann constant). The fitting shown in the inset to Fig.~\ref{SdH_analysis}b yielded $m^*_\alpha=\,0.11m_e$ and $m^*_\beta=\,0.21m_e$. It is worth noting that the effective mass determined for the $\alpha$ Fermi surface pocket in PtTe$_2$ is almost equal to that reported for the Pd-bearing counterpart.\cite{Wang2016i}

\begin{figure}[h]
	\centering
	\includegraphics[width=17cm]{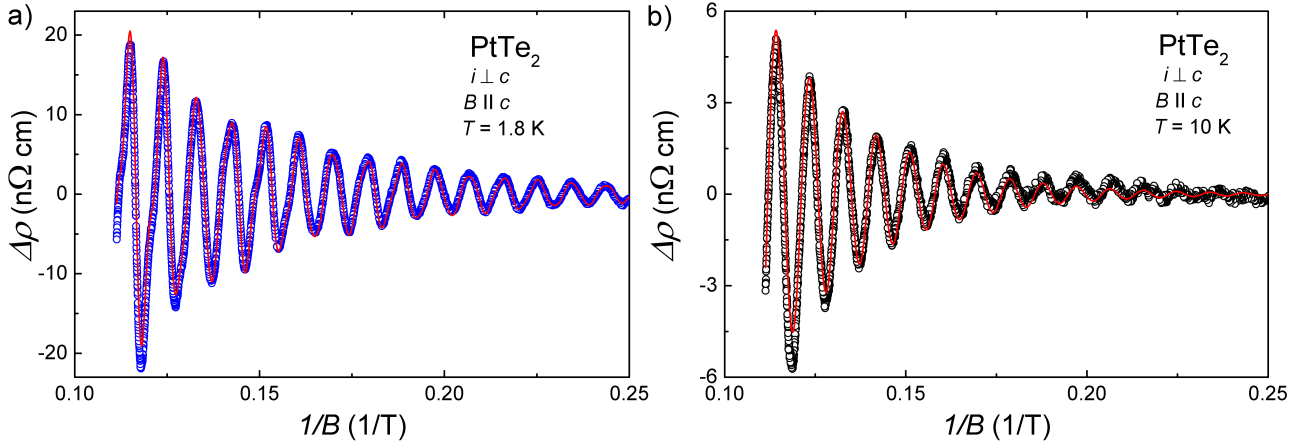}
	\caption{\textbf{Lifshitz-Kosevich analysis of the magnetotransport in PtTe$_2$.} Oscillating part of the electrical resistivity measured at (a) $T=1.8$\,K and (b) $T=10$\,K, plotted as a function of inverse magnetic field. Solid red lines correspond to the results of fitting Eq.~\ref{full_LK_eq} to the experimental data.
	\label{SdH_LK_fit}}
\end{figure}

In the next step, we attempted to determine the phase shift, $\varphi_i$, in the SdH oscillations, which is directly related to the Berry phase, $\varphi_{\rm{B},i}$, of the carries involved. There are known a few methods of extracting $\varphi_i$, and the proper interpretation of their values remains debatable.\cite{Ando2013,Wang2016k,Li2018d,Taskin2011b} The most reliable approach is direct fitting to the experimental data of the Lifshitz-Kosevich (LK) function:\cite{Shoenberg1984}
\begin{equation}
	\Delta\rho \propto \frac{1}{\sqrt{B}}\sum_{i}\frac{p_i\lambda m^*_iT/B}{\sinh(p_i\lambda m^*_iT/B)}\exp(-p_i\lambda m^*_iT_{\rm{D,i}}/B)\cos\bigg(2\pi\big(p_if_i/B+\varphi_i\big)\bigg),
	\label{full_LK_eq}
\end{equation}
where $p_i$ is the harmonic number, $f_i$ is the oscillation frequency, and $T_{\rm{D,i}}$ stands for the Dingle temperature. In Fig.~\ref{SdH_LK_fit}a, there is shown the result of fitting the LK function to the oscillating resistivity of PtTe$_2$ observed at $T=1.8$\,K (note the red solid line). As discussed above, at this temperature, the FFT spectrum comprises three peaks, and thus the sum in Eq.~\ref{full_LK_eq} consists of as many as three contributions. In order to reduce the total number of free parameters in this equation, the effective masses were fixed at the values obtained from Eq.~\ref{LK_eq} (see above). With this simplification, one obtained the parameters: $f_\alpha$ = 108.1~T, $T_{\rm{D},\alpha}=9.1$\,K and $\varphi_\alpha=0.65$ for the $\alpha$ band, and $f_\beta$ = 246~T, $T_{\rm{D},\beta}=5$\,K and $\varphi_\beta=0.54$ for the $\beta$ band. Remarkably, the so-obtained values of $f_i$ are almost identical to those derived from the afore-described FFT analysis, hence confirming the internal consistency of the approach applied. Using the Dingle temperatures, the quantum relaxation time, $\tau_{q,i}$, could be calculated from the relation $\tau_{q,i}=\hbar/(2\pi k_{\rm B}T_{\rm{D},i})$ to be equal to $1.34\times10^{-13}$\,s and $2.43\times10^{-13}$\,s for the $\alpha$ and $\beta$ bands, respectively. Then, the quantum mobility of charge carriers, $\mu_{q,i}$, were estimated from the relationship $\mu_{q,i}=e\tau_{q,i}/m^*_i$ to be $2138\,\rm{cm^2V^{-1}s^{-1}}$ and $2038\,\rm{cm^2V^{-1}s^{-1}}$ for the $\alpha$ and $\beta$ bands, respectively.

The phase shift $\varphi_i$ in Eq.~\ref{full_LK_eq} is generally a sum $\varphi_i=-1/2+\varphi_{B,i}+\delta_i$, where $\delta_i$ represents the dimension-dependent correction to the phase shift.\cite{Li2018d} In two-dimensional (2D) case, this parameter amounts zero, while in three-dimensional (3D) case $\delta_i$ is equal to $\pm1/8$, and its sign depends on type of charge carriers and kind of cross-section extremum. Supposing that the SdH oscillations in PtTe$_2$ originate from 3D bands with carriers moving on their maximal orbits, one can set $\delta=-1/8$ for electrons and $\delta=1/8$ for holes. With this assumption, the Berry phases $\varphi_{B,\alpha}=0.55\pi$ and $\varphi_{B,\beta}=1.83\pi$ were obtained for the $\alpha$ and $\beta$ bands, respectively.

To check the reliability of the LK analysis performed, Eq.~\ref{full_LK_eq} was also used to describe the experimental data measured at $T=10$\,K. At this temperature, just one peak in the FFT spectrum is discernible (see Fig.~\ref{SdH_analysis}b), which corresponds to the $\alpha$ Fermi surface pocket. The result of fitting the LK formula is presented in Fig.~\ref{SdH_LK_fit}b (note the red solid line), and the so-derived values of the parameters are: $f_\alpha$ = 108.1~T, $T_{\rm{D},\alpha}=12.6$\,K and $\varphi_{B,\alpha}=0.46\pi$. Notably, the agreement between the $f_\alpha$ values obtained at $T=10$\,K and $T=1.8$\,K is perfect. The value of $T_{\rm{D},\alpha}$ implies $\mu_q=1544\,\rm{cm^2V^{-1}s^{-1}}$. Clearly, with increasing temperature, the Dingle temperature increases and consequently the quantum charge carriers mobility becomes smaller, which is probably due to increasing the scattering rate. In turn, the Berry phase of the $\alpha$ band was found almost independent of temperature. All the parameters obtained from the LK approach to the magnetotransport in PtTe$_2$ are gathered in Table~\ref{L_K_param}.

\begin{table}[h]
\centering
	\caption{\textbf{Parameters extracted from the LK analysis of the SdH oscillations in PtTe$_2$}. $T$ - temperature; $f_{i}^{\rm{FFT}}$ - oscillation frequency obtained from the FFT analysis; $f_{i}$ - oscillation frequency obtained from Eq.~\ref{full_LK_eq}; $m^*$ - effective mass; $T_{\rm{D}}$ - Dingle temperature; $\tau_q$ - quantum relaxation time; $\mu_q$ - quantum mobility; $\varphi_{\rm{B}}$ - Berry phase.}
	\begin{tabular*}{0.85\textwidth}{@{\extracolsep{\fill}}*{9}{c}} \hline\hline
		$T$ & band & $f_{i}^{\rm{FFT}}$ & $f_i$ & $m^*$ & $T_{\rm{D}}$ & $\tau_q$ & $\mu_q$ & $\varphi_{\rm{B}}$ \\
		(K) & & (T) & (T) & ($m_e$) & (K) & (s) & ($\rm{cm^2V^{-1}s^{-1}}$) & \\\hline
		\multirow{2}{*}{1.8} & $\alpha$ & 108 & 108.1 & 0.11 & 9.1 & $1.34\times10^{-13}$ & 2138 & $0.55\pi$ \\
		& $\beta$ & 246 & 246 & 0.21 & 5 & $2.43\times10^{-13}$ & 2038 & $1.83\pi$ \\
		10 & $\alpha$ & 108 & 108.5 & 0.11 & 12.6 & $9.65\times10^{-14}$ & 1544 & $0.46\pi$ \\
		\hline\hline
	\end{tabular*}\label{LMtable}
	\label{L_K_param}
\end{table}

\begin{figure}[h]
	\centering
	\includegraphics[width=8.5cm]{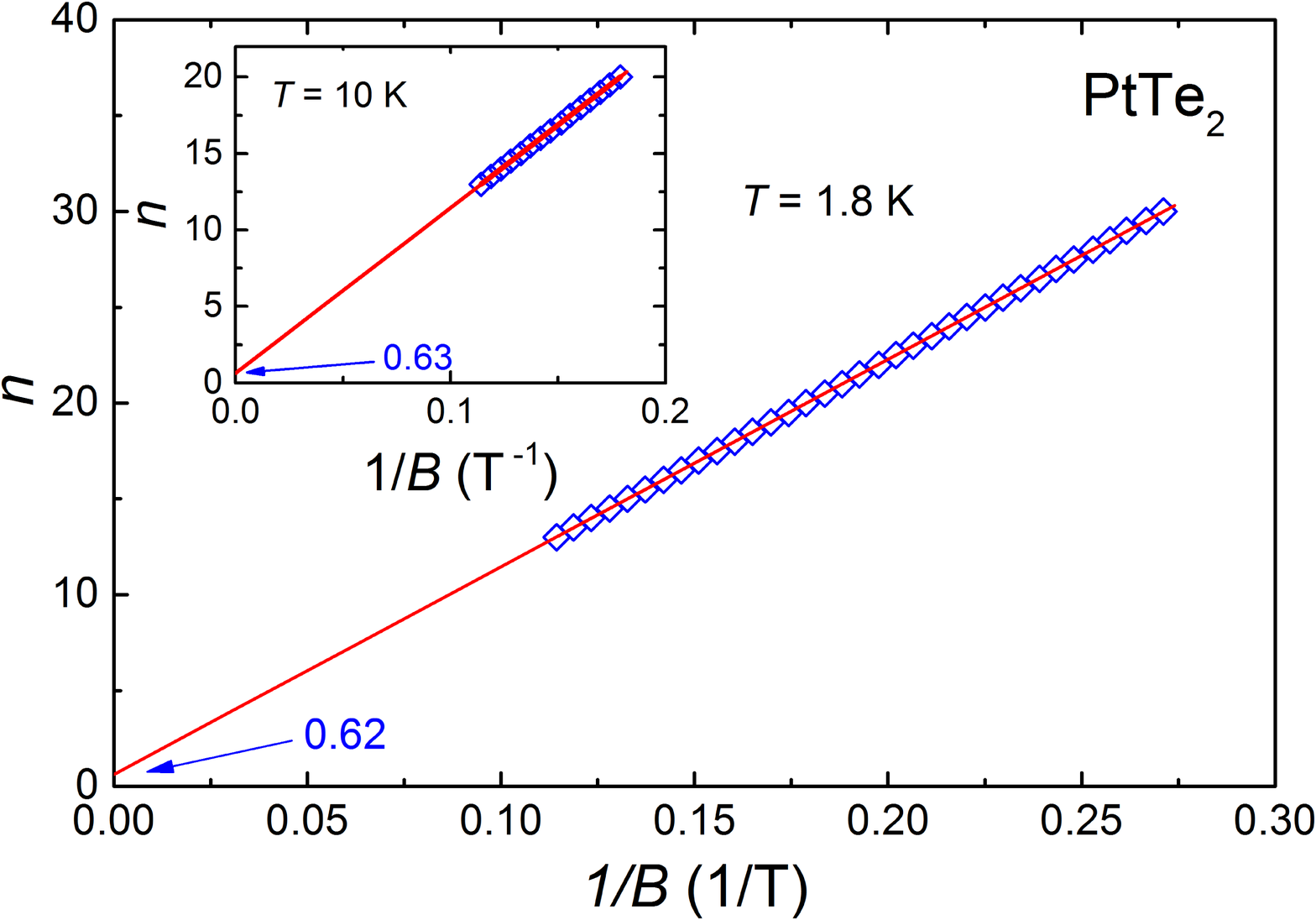}
	\caption{\textbf{Landau level fan diagrams for PtTe$_2$.} The Landau level indices, $n$, determined for the $\alpha$ Fermi surface pocket at $T=1.8$\,K plotted as a function of inverse magnetic field. Inset shows the LL plot of the SdH oscillations measured at $T=10$\,K. Solid lines represent the linear fits to the LL data.
		\label{Berry}}
\end{figure}

Another commonly applied technique for Berry phase derivation is using Landau level (LL) fan diagrams. Though in case of multi-frequency oscillations this method is obstructed by possible superposition of the quantum oscillation peaks that hinders precise determination of the Landau level index for a given frequency,\cite{Hu2016} we made an attempt to construct the LL fan diagram for the $\alpha$ Fermi pocket in PtTe$_2$. As it is apparent from Fig.~\ref{SdH_analysis}b, the FFT maximum occurring at $f_{\alpha}^{FFT}=108$\,T is fairly well separated from the other FFT peaks, and hence one could filter this oscillation with reasonably high accuracy. For PtTe$_2$ one finds $\rho > \rho_{xy}$ ($\rho_{xy}$ is the Hall resistivity discussed in the next section), and therefore the maxima in the oscillatory resistivity measured at $T=1.8$\,K (see Fig.~\ref{SdH_LK_fit}a) were numbered by integers, $n$, and the minima by half-integers, $n+1/2$. The result of this approach is shown in the main panel of Fig.~\ref{Berry}. A linear fit of the LL indices (note the solid line) gives an intercept of 0.62, which corresponds to the Berry phase $\varphi_{B,\alpha}=0.51\pi$. In turn, the slope of this straight line defines the oscillation frequency $f_\alpha= 108.3$~T.

At $T=10$\,K, the electrical resistivity of PtTe$_2$ oscillates in the transverse magnetic field with only one frequency (cf. Fig.~\ref{SdH_analysis}b), so building the LL fan diagram is straightforward. As can be inferred from the inset to Fig.~\ref{Berry}, the LL indices plot yields the intercept 0.63, which is almost the same as that obtained at the lower temperature. Also, the slope of the straight line ($f_\alpha= 108.3$~T) is identical with that determined at 1.8\,K, and furthermore it is very close to the FFT value (see Table~\ref{L_K_param}). Most importantly, all the parameters extracted from the LL indices plots are in perfect agreement with the quantities obtained for the $\alpha$ Fermi pocket from the LK analysis, which unambiguously corroborates the correctness of both techniques applied for PtTe$_2$.

\subsection*{Hall effect}

\begin{figure}[h]
	\centering
	\includegraphics[width=17cm]{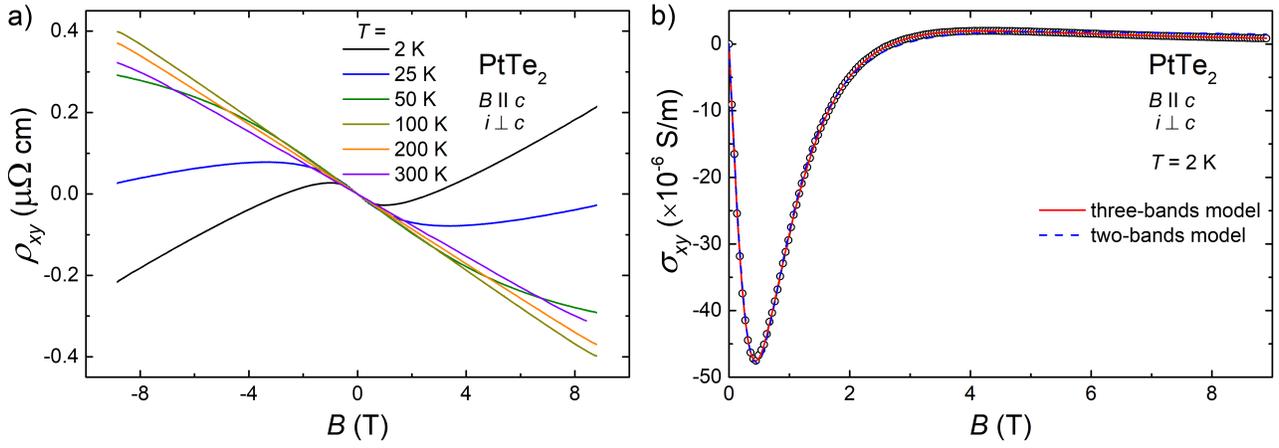}
	\caption{\textbf{Hall effect in PtTe$_2$}. (a) Magnetic field dependencies of the Hall resistivity measured at several different temperatures with $i \perp c$ axis and $B \parallel c$ axis. (b) Hall conductivity as a function of magnetic field at $T=2$\,K. Blue dashed line represents the fit with two-bands model, and red solid line stands for result obtained with three-bands model.
		\label{Hall}}
\end{figure}

The results of Hall resistivity measurements, performed on single-crystalline PtTe$_2$ with electric current flowing within the basal plane of the hexagonal unit cell and magnetic field applied along the $c$ axis, are shown in Fig.~\ref{Hall}a. At $T=$ 2~K, $\rho_{xy}(B)$ behaves in a rather complex manner. In weak magnetic fields, it is negative and exhibits a shallow minimum. Near 2~T, the Hall resistivity changes sign to positive, and then its magnitude increases with increasing $B$. The $\rho_{xy}(B)$ isotherm measured at 25~K shows a fairly similar field variation, yet the positive contribution in strong fields remains too small to cause sign reversal. At higher temperatures, one observes a gradual straightening of the $\rho_{xy}(B)$ with rising $T$. The overall behavior of the Hall effect in PtTe$_2$ confirms the multi-band character of the electrical transport in this material. It is worth recalling that very similar Hall response was observed for the closely related compound PdTe$_2$.\cite{Fei2017}

For quantitative analysis of the experimental data, first a two-bands Drude model was applied. For this purpose, $\rho_{xy}(B)$ measured at $T=2$\,K was converted to the Hall conductivity $\sigma_{xy}=\rho_{xy}/(\rho_{xy}^2+\rho^2)$, as displayed in Figure~\ref{Hall}b. Next, $\sigma_{xy}(B)$ was fitted by the formula:

\begin{equation}
	\sigma_{xy}(B)=eB\Bigg(\frac{n_{h} \mu_{h}^2}{1+(\mu_hB)^2}+\frac{n_{e1} \mu_{e1}^2}{1+(\mu_{e1}B)^2}\Bigg),
	\label{two-band-sigma-model}
\end{equation}
where $n_{e1}$ and $\mu_{e1}$, $n_h$ and $\mu_h$ stand for the carrier concentrations and the carrier mobilities of electron- and hole-like bands, respectively. As can be inferred from Fig.~\ref{Hall}b, the so-obtained approximation of the measured $\sigma_{xy}(B)$ data (note the blue dashed line) is not ideal. An obvious reason for the discrepancy between the experiment and the two-band model could be contribution from another band, the presence of which was revealed in the ab-initio calculations of the electronic structure of PtTe$_2$.\cite{Zhang2017e,Yan2017a}

Therefore, in the next step, the Hall conductivity was analysed in terms of a three-bands model:

\begin{equation}
	\sigma_{xy}(B)=eB\Bigg(\frac{n_{h} \mu_{h}^2}{1+(\mu_hB)^2}+\frac{n_{e1} \mu_{e1}^2}{1+(\mu_{e1}B)^2}+\frac{n_{e2} \mu_{e2}^2}{1+(\mu_{e2}B)^2}\Bigg).
	\label{three-band-sigma-model}
\end{equation}
where $n_{e2}$ and $\mu_{e2}$ account for the carrier concentration and the carrier mobility, respectively, of another electron-like band in PtBi$_2$. The result of fitting Eq.~\ref{three-band-sigma-model} to the experimental $\sigma_{xy}(B)$ data is shown as red solid line in Fig.~\ref{Hall}b. Clearly, the obtained description is much better than that with the two-bands model.

\begin{table}[h]
	\centering
	\caption{\textbf{Parameters obtained for PtTe$_2$ from multi-bands models analyses of the the Hall effect data.} $n_h$ - hole concentration; $\mu_h$ - hole mobility; $n_{e1}$ and $n_{e2}$ - electron concentrations; $\mu_{e1}$ and $\mu_{e2}$ - electron mobilities.}
	\begin{tabular*}{0.85\textwidth}{@{\extracolsep{\fill}}*{7}{c}} \hline\hline
		Model & $n_h$ & $\mu_h$& $n_{e1}$ & $\mu_{e1}$ & $n_{e2}$ & $\mu_{e2}$ \\
		& (cm$^{-3}$) & ($\rm{cm^2V^{-1}s^{-1}}$) & (cm$^{-3}$) & ($\rm{cm^2V^{-1}s^{-1}}$) & (cm$^{-3}$) & ($\rm{cm^2V^{-1}s^{-1}}$) \\\hline
		two-bands & $5.99\!\times\!10^{20}$ & 7179 & $5.05\!\times\!10^{20}$ & 17373 & - & -\\
		three-bands & $7.91\!\times\!10^{20}$ & 4740 & $3.8\!\times\!10^{20}$ & 19240 & $3.82\!\times\!10^{20}$ & 2564 \\
		\hline\hline
	\end{tabular*}\label{LMtable}
	\label{Hall_parameters}
\end{table}

The fitting parameters derived in the two approaches are listed in Table~\ref{Hall_parameters}. Both models yielded large carriers concentrations of the order of $10^{20}\,\rm{cm^{-3}}$. It is worth noting that very similar charge densities were found in the Dirac semimetal PtBi$_2$.\cite{Gao2017} On the contrary, for the type-II Weyl semimetal WTe$_2$ the carrier concentrations were reported to be up to two orders of magnitude larger than those in PtTe$_2$.\cite{Wang2017b} As regards the level of carrier compensation, the two-bands model yielded considerable charge imbalance given by the ratio $n_h/n_{e1}=1.19$, however the three-bands model led to fairly balanced scenario $n_h/(n_{e1}+n_{e2})=1.04$. Recently, similar degree of electron-hole compensation was established, e.g., in semimetallic monobismuthides YBi and LuBi.\cite{Pavlosiuk2017b} The mobilities of charge carriers in PtTe$_2$ were found very high, especially that obtained for one of the electron-like Fermi surface pockets ($\mu_{e1} \sim2\!\times\!10^4\,\rm{cm^2V^{-1}s^{-1}}$). Though the latter value is not such large as the carriers mobilities in Cd$_3$As$_2$ (Ref.\cite{Liang2014}) or NbP (Ref.\cite{Shekhar2015b}), it exceeds the values reported for type-II Weyl semimetals MoTe$_2$,\cite{Chen2016a} WTe$_2$,\cite{Luo2015a} and WP$_2$.\cite{Wang2017b} It is worth noting that the carriers mobility derived from the Hall effect data are larger than the quantum mobilities determined in the analyses of the SdH oscillations. Similar finding was reported for other TSs, like Cd$_3$As$_2$,\cite{Liang2014} ZrSiS,\cite{Hu2017b} WP$_2$,\cite{Kumar2017c} and PtBi$_2$.\cite{Gao2017}
The discrepancy likely arises due to the fact that the quantum mobility is affected by all possible scattering processes, whereas the Hall mobility is sensitive to small-angle scattering only.\cite{Shoenberg1984}

\subsection*{Angle-dependent magnetoresistance}

\begin{figure}[h]
	\centering
	\includegraphics[width=8.5cm]{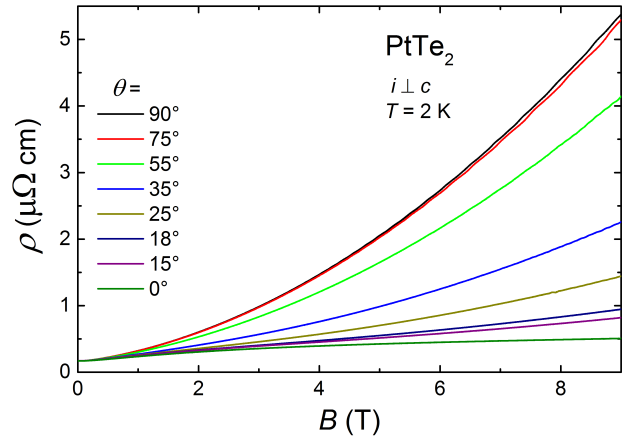}
	\caption{\textbf{Angle-dependent magnetotransport in PtTe$_2$}. Magnetic field dependencies of the electrical resistivity measured at $T$ = 2~K with $i \perp c$ axis and external magnetic field oriented at several different angles in relation to the electric current direction.	
		\label{rho_rotator}}
\end{figure}

In order to check whether PtTe$_2$ demonstrates CMA, angle-dependent magnetotransport measurements were performed at $T$ = 2~K. In these experiments, electric current was always flowing within the hexagonal $a-b$ plane, while the angle $\theta$ between current and magnetic field direction was varied from $\theta=90^{\circ}$ ($B \perp i$) to $\theta=0^{\circ}$ ($B \parallel i$). As can be inferred from Fig.~\ref{rho_rotator}, the electrical resistivity rapidly decreases on deviating from the transverse configuration, and eventually for the longitudinal geometry $\rho$ measured in $B$ = 9~T is about an order of magnitude smaller than that for $B \perp i$.

Clearly, the longitudinal MR experiments did not provide any evidence for CMA in PtTe$_2$. A possible source for that may be large contribution of non-Dirac states to the measured resistivity. At odds with the Drude theory, which predicts zero MR for $B \parallel i$,\cite{Pippard2009} in several materials sizeable positive longitudinal MR was observed.
Among the theories which interpret this phenomenon,\cite{Miller1996,Argyres1956,Stroud1976,Pal2010} that accounting for Fermi surface anisotropy\cite{Pal2010} seems appropriate for PtTe$_2$. Within the latter approach, positive longitudinal MR up to $\sim100\%$ can be expected for strongly anisotropic systems. In consequence, CMA would be discernible only if its negative contribution to the longitudinal MR is larger than the positive term due to trivial electronic bands.

\section*{Conclusions}

Our comprehensive investigations of the galvanomagnetic properties of the alleged type-II Dirac semimetal PtTe$_2$, performed on high-quality single crystals, have not provided any definitive proof of the the presence of Dirac states in this material. The conclusion was hampered by the existence of trivial bands at the Fermi level, which significantly contribute to the electrical transport. In particular, CMA effect was not resolved, and transverse MR was found to obey the Kohler's scaling. From the analysis of the Hall effect and the SdH oscillations, very high mobilities of charge carriers with small effective masses were extracted. However, the derived Berry phases, different from the value of $\pi$ expected for Dirac fermions, indicate that the SdH effect is governed predominantly by trivial electronic states. This finding is in concert with the electronic band structure calculations, which showed that the Dirac point in PtTe$_2$ is located below the Fermi level.\cite{Yan2017a} Further investigations performed on suitably doped or pressurized material might result in observation of clear contribution of Dirac states to its transport properties, caused by appropriate tuning the chemical potential. Based on the hitherto obtained results, PtTe$_2$ can be classified as a semimetal with moderate degree of the charge carriers compensation.

\section*{Methods}

Single crystals of PtTe$_2$ were grown by flux method. High-purity constituents (Pt 5N, Te 6N), taken in atomic ratio 1:20, were placed in an alumina crucible covered by molybdenum foil strainer and capped with another inverted alumina crucible. This set was sealed inside a quartz tube under partial Ar gas atmosphere. The ampoule tube was heated up to 1150$^{\circ}$C, held at this temperature for 24 hours, then quickly cooled down to 850$^{\circ}$C at a rate of 50$^{\circ}$C/h, kept at this temperature for 360 hours, followed by slow cooling down to 550$^{\circ}$C at a rate of 5$^{\circ}$C/h. Subsequently, the tube was quenched in cold water. Upon flux removal by centrifugation, multitude of single crystals with typical dimensions $3\times\!2\!\times\!0.4$\,mm$^3$ were isolated. Their had metallic luster and were found stable against air and moisture.

Chemical composition of the single crystals obtained was checked by energy-dispersive X-ray analysis using a FEI scanning electron microscope equipped with an EDAX Genesis XM4 spectrometer. The average elemental ratio Pt : Te = 35.2(5) : 64.8(3) was derived, in accord with the expected stoichiometry. The crystal structure of the single crystals was examined by X-ray diffraction on a KUMA Diffraction KM-4 four-circle diffractometer equipped with a CCD camera, using graphite-monochromatized Cu-K$\alpha 1$ radiation. The hexagonal CdI$_2$-type crystal structure (space group $P\overline{3}m1$, Wyckoff No. 164) reported in Ref.\cite{Faruseth1965} was confirmed, with the lattice parameters very close to the literature values.

Crystallinity and orientation of the crystal used in the electrical transport studies was checked by Laue backscattering technique employing a Proto LAUE-COS system. Due to the layered crystal structure of PtTe$_2$ it was possible to obtain very thin samples by scotch-tape technique with their surface corresponding to the $a-b$ plane of the hexagonal unit cell of the compound. Rectangular-shaped specimen with dimensions $2.9\times\!1.3\!\times\!0.04$\,mm$^3$ was cut from the cleaved single crystal using a scalpel. Electrical contacts were made from $50\,\mu$m thick silver wires attached to the sample using silver epoxy paste. Electrical transport measurements were carried out within the temperature range $2-300$\,K and in magnetic field up to 9\,T using a conventional four-point ac technique implemented in a Quantum Design PPMS platform.

\section*{Author contributions statement} D.K. conceived the experiments and performed preliminary electrical transport studies. O.P. conducted the experiments and analysed the data. Both authors contributed to discussion of the results and writing the manuscript.


\begin{thebibliography}{10}
	\expandafter\ifx\csname url\endcsname\relax
	\def\url#1{\texttt{#1}}\fi
	\expandafter\ifx\csname urlprefix\endcsname\relax\def\urlprefix{URL }\fi
	\expandafter\ifx\csname doiprefix\endcsname\relax\def\doiprefix{DOI }\fi
	\providecommand{\bibinfo}[2]{#2}
	\providecommand{\eprint}[2][]{\url{#2}}
	
	\bibitem{Hasan2017}
	\bibinfo{author}{Hasan, M.~Z.}, \bibinfo{author}{Xu, S.-Y.},
	\bibinfo{author}{Belopolski, I.} \& \bibinfo{author}{Huang, S.-M.}
	\newblock \bibinfo{journal}{\bibinfo{title}{{Discovery of Weyl fermion
				semimetals and topological Fermi arc states}}}.
	\newblock {\emph{\JournalTitle{Ann. Rev. Cond. Mat. Phys.}}}
	\textbf{\bibinfo{volume}{8}}, \bibinfo{pages}{289--309}
	(\bibinfo{year}{2017}).
	
	\bibitem{Armitage2017a}
	\bibinfo{author}{Armitage, N.~P.}, \bibinfo{author}{Mele, E.~J.} \&
	\bibinfo{author}{Vishwanath, A.}
	\newblock \bibinfo{journal}{\bibinfo{title}{{Weyl and Dirac semimetals in three
				dimensional solids}}}.
	\newblock {\emph{\JournalTitle{Rev. Mod. Phys.}}}
	\textbf{\bibinfo{volume}{90}}, \bibinfo{pages}{015001}
	(\bibinfo{year}{2017}).
	
	\bibitem{Yan2017a}
	\bibinfo{author}{Yan, M.} \emph{et~al.}
	\newblock \bibinfo{journal}{\bibinfo{title}{{Lorentz-violating type-II Dirac
				fermions in transition metal dichalcogenide PtTe$_2$}}}.
	\newblock {\emph{\JournalTitle{Nat. Commun.}}} \textbf{\bibinfo{volume}{8}},
	\bibinfo{pages}{257} (\bibinfo{year}{2017}).
	
	\bibitem{Soluyanov2015a}
	\bibinfo{author}{Soluyanov, A.~A.} \emph{et~al.}
	\newblock \bibinfo{journal}{\bibinfo{title}{{Type-II Weyl semimetals}}}.
	\newblock {\emph{\JournalTitle{Nature}}} \textbf{\bibinfo{volume}{527}},
	\bibinfo{pages}{495--498} (\bibinfo{year}{2015}).
	
	\bibitem{Yang2014a}
	\bibinfo{author}{Yang, B.-J.} \& \bibinfo{author}{Nagaosa, N.}
	\newblock \bibinfo{journal}{\bibinfo{title}{{Classification of stable
				three-dimensional Dirac semimetals with nontrivial topology}}}.
	\newblock {\emph{\JournalTitle{Nat. Commun.}}} \textbf{\bibinfo{volume}{5}},
	\bibinfo{pages}{4898} (\bibinfo{year}{2014}).
	
	\bibitem{Nielsen1983}
	\bibinfo{author}{Nielsen, H.~B.} \& \bibinfo{author}{Ninomiya, M.}
	\newblock \bibinfo{journal}{\bibinfo{title}{{The Adler-Bell-Jackiw anomaly and
				Weyl fermions in a crystal}}}.
	\newblock {\emph{\JournalTitle{Phys. Lett. B}}} \textbf{\bibinfo{volume}{130}},
	\bibinfo{pages}{389--396} (\bibinfo{year}{1983}).
	
	\bibitem{Li2015c}
	\bibinfo{author}{Li, C.~Z.} \emph{et~al.}
	\newblock \bibinfo{journal}{\bibinfo{title}{{Giant negative magnetoresistance
				induced by the chiral anomaly in individual Cd$_3$As$_2$ nanowires}}}.
	\newblock {\emph{\JournalTitle{Nature Communications}}}
	\textbf{\bibinfo{volume}{6}}, \bibinfo{pages}{10137} (\bibinfo{year}{2015}).
	
	\bibitem{Li2016}
	\bibinfo{author}{Li, Q.} \emph{et~al.}
	\newblock \bibinfo{journal}{\bibinfo{title}{{Chiral magnetic effect in
				ZrTe$_5$}}}.
	\newblock {\emph{\JournalTitle{Nature Physics}}} \textbf{\bibinfo{volume}{12}},
	\bibinfo{pages}{550} (\bibinfo{year}{2016}).
	
	\bibitem{Hirschberger2016a}
	\bibinfo{author}{Hirschberger, M.} \emph{et~al.}
	\newblock \bibinfo{journal}{\bibinfo{title}{{The chiral anomaly and thermopower
				of Weyl fermions in the half-Heusler GdPtBi}}}.
	\newblock {\emph{\JournalTitle{Nature Materials}}}
	\textbf{\bibinfo{volume}{15}}, \bibinfo{pages}{1161} (\bibinfo{year}{2016}).
	
	\bibitem{Niemann2016a}
	\bibinfo{author}{Niemann, A.~C.} \emph{et~al.}
	\newblock \bibinfo{journal}{\bibinfo{title}{{Chiral magnetoresistance in the
				Weyl semimetal NbP}}}.
	\newblock {\emph{\JournalTitle{Sci. Rep.}}} \textbf{\bibinfo{volume}{7}},
	\bibinfo{pages}{43394} (\bibinfo{year}{2017}).
	
	\bibitem{Xiong2015}
	\bibinfo{author}{Xiong, J.} \emph{et~al.}
	\newblock \bibinfo{journal}{\bibinfo{title}{{Evidence for the chiral anomaly in
				the Dirac semimetal Na$_3$Bi}}}.
	\newblock {\emph{\JournalTitle{Science}}} \textbf{\bibinfo{volume}{350}},
	\bibinfo{pages}{413} (\bibinfo{year}{2015}).
	
	\bibitem{Lv2017}
	\bibinfo{author}{Lv, Y.-Y.} \emph{et~al.}
	\newblock \bibinfo{journal}{\bibinfo{title}{{Experimental observation of
				anisotropic Adler-Bell-Jackiw anomaly in type-II Weyl semimetal WTe$_{1.98}$
				crystals at the quasiclassical regime}}}.
	\newblock {\emph{\JournalTitle{Physical Review Letters}}}
	\textbf{\bibinfo{volume}{118}}, \bibinfo{pages}{096603}
	(\bibinfo{year}{2017}).
	
	\bibitem{Wang2016m}
	\bibinfo{author}{Wang, L.-X.}, \bibinfo{author}{Li, C.-Z.},
	\bibinfo{author}{Yu, D.-P.} \& \bibinfo{author}{Liao, Z.-M.}
	\newblock \bibinfo{journal}{\bibinfo{title}{{Aharonov–Bohm oscillations in
				Dirac semimetal Cd$_3$As$_2$ nanowires}}}.
	\newblock {\emph{\JournalTitle{Nature Communications}}}
	\textbf{\bibinfo{volume}{7}}, \bibinfo{pages}{10769} (\bibinfo{year}{2016}).
	
	\bibitem{Deng2016}
	\bibinfo{author}{Deng, K.} \emph{et~al.}
	\newblock \bibinfo{journal}{\bibinfo{title}{{Experimental observation of
				topological Fermi arcs in type-II Weyl semimetal MoTe$_2$}}}.
	\newblock {\emph{\JournalTitle{Nat. Phys.}}} \textbf{\bibinfo{volume}{12}},
	\bibinfo{pages}{1105--1110} (\bibinfo{year}{2016}).
	
	\bibitem{Li2017b}
	\bibinfo{author}{Li, P.} \emph{et~al.}
	\newblock \bibinfo{journal}{\bibinfo{title}{{Evidence for topological type-II
				Weyl semimetal WTe$_2$}}}.
	\newblock {\emph{\JournalTitle{Nature Communications}}}
	\textbf{\bibinfo{volume}{8}}, \bibinfo{pages}{2150} (\bibinfo{year}{2017}).
	
	\bibitem{Lv2015}
	\bibinfo{author}{Lv, B.~Q.} \emph{et~al.}
	\newblock \bibinfo{journal}{\bibinfo{title}{{Experimental discovery of weyl
				semimetal TaAs}}}.
	\newblock {\emph{\JournalTitle{Physical Review X}}}
	\textbf{\bibinfo{volume}{5}}, \bibinfo{pages}{031013} (\bibinfo{year}{2015}).
	
	\bibitem{Chang2017a}
	\bibinfo{author}{Chang, T.-R.} \emph{et~al.}
	\newblock \bibinfo{journal}{\bibinfo{title}{{Type-II symmetry-protected
				topological Dirac semimetals}}}.
	\newblock {\emph{\JournalTitle{Phys. Rev. Lett.}}}
	\textbf{\bibinfo{volume}{119}}, \bibinfo{pages}{026404}
	(\bibinfo{year}{2017}).
	
	\bibitem{Zhang2017e}
	\bibinfo{author}{Zhang, K.} \emph{et~al.}
	\newblock \bibinfo{journal}{\bibinfo{title}{{Experimental evidence for type-II
				Dirac semimetal in PtSe$_2$}}}.
	\newblock {\emph{\JournalTitle{Phys. Rev. B}}} \textbf{\bibinfo{volume}{96}},
	\bibinfo{pages}{125102} (\bibinfo{year}{2017}).
	
	\bibitem{Autes2016}
	\bibinfo{author}{Aut{\`{e}}s, G.}, \bibinfo{author}{Gresch, D.},
	\bibinfo{author}{Troyer, M.}, \bibinfo{author}{Soluyanov, A.~A.} \&
	\bibinfo{author}{Yazyev, O.~V.}
	\newblock \bibinfo{journal}{\bibinfo{title}{{Robust type-II Weyl semimetal
				phase in transition metal diphosphides $X$P$_2$ ($X$=Mo, W)}}}.
	\newblock {\emph{\JournalTitle{Phys. Rev. Lett.}}}
	\textbf{\bibinfo{volume}{117}}, \bibinfo{pages}{066402}
	(\bibinfo{year}{2016}).
	
	\bibitem{Huang2016b}
	\bibinfo{author}{Huang, H.}, \bibinfo{author}{Zhou, S.} \&
	\bibinfo{author}{Duan, W.}
	\newblock \bibinfo{journal}{\bibinfo{title}{{Type-II Dirac fermions in the
				PtSe$_2$ class of transition metal dichalcogenides}}}.
	\newblock {\emph{\JournalTitle{Phys. Rev. B}}} \textbf{\bibinfo{volume}{94}},
	\bibinfo{pages}{121117} (\bibinfo{year}{2016}).
	
	\bibitem{Bruno2016}
	\bibinfo{author}{Bruno, F.~Y.} \emph{et~al.}
	\newblock \bibinfo{journal}{\bibinfo{title}{{Observation of large topologically
				trivial Fermi arcs in the candidate type-II Weyl semimetal WTe$_2$}}}.
	\newblock {\emph{\JournalTitle{Phys. Rev. B}}} \textbf{\bibinfo{volume}{94}},
	\bibinfo{pages}{121112} (\bibinfo{year}{2016}).
	
	\bibitem{Noh2017}
	\bibinfo{author}{Noh, H.-J.} \emph{et~al.}
	\newblock \bibinfo{journal}{\bibinfo{title}{{Experimental realization of
				type-II Dirac fermions in a PdTe$_2$ Superconductor}}}.
	\newblock {\emph{\JournalTitle{Phys. Rev. Lett.}}}
	\textbf{\bibinfo{volume}{119}}, \bibinfo{pages}{016401}
	(\bibinfo{year}{2017}).
	
	\bibitem{Wang2016i}
	\bibinfo{author}{Wang, Y.} \emph{et~al.}
	\newblock \bibinfo{journal}{\bibinfo{title}{{De Hass-van Alphen and
				magnetoresistance reveal predominantly single-band transport behavior in
				PdTe$_2$}}}.
	\newblock {\emph{\JournalTitle{Sci. Rep.}}} \textbf{\bibinfo{volume}{6}},
	\bibinfo{pages}{31554} (\bibinfo{year}{2016}).
	
	\bibitem{Fei2017}
	\bibinfo{author}{Fei, F.} \emph{et~al.}
	\newblock \bibinfo{journal}{\bibinfo{title}{{Nontrivial Berry phase and type-II
				Dirac transport in the layered material PdTe$_2$}}}.
	\newblock {\emph{\JournalTitle{Phys. Rev. B}}} \textbf{\bibinfo{volume}{96}},
	\bibinfo{pages}{041201} (\bibinfo{year}{2017}).
	
	\bibitem{Sun2016a}
	\bibinfo{author}{Sun, S.}, \bibinfo{author}{Wang, Q.}, \bibinfo{author}{Guo,
		P.-J.}, \bibinfo{author}{Liu, K.} \& \bibinfo{author}{Lei, H.}
	\newblock \bibinfo{journal}{\bibinfo{title}{{Large magnetoresistance in LaBi:
				Origin of field-induced resistivity upturn and plateau in compensated
				semimetals}}}.
	\newblock {\emph{\JournalTitle{New J. Phys.}}} \textbf{\bibinfo{volume}{18}},
	\bibinfo{pages}{082002} (\bibinfo{year}{2016}).
	
	\bibitem{Pavlosiuk2017}
	\bibinfo{author}{Pavlosiuk, O.}, \bibinfo{author}{Kleinert, M.},
	\bibinfo{author}{Swatek, P.}, \bibinfo{author}{Kaczorowski, D.} \&
	\bibinfo{author}{Wi{\'{s}}niewski, P.}
	\newblock \bibinfo{journal}{\bibinfo{title}{{Fermi surface topology and
				magnetotransport in semimetallic LuSb}}}.
	\newblock {\emph{\JournalTitle{Sci. Rep.}}} \textbf{\bibinfo{volume}{7}},
	\bibinfo{pages}{12822} (\bibinfo{year}{2017}).
	
	\bibitem{Pavlosiuk2017b}
	\bibinfo{author}{Pavlosiuk, O.}, \bibinfo{author}{Swatek, P.},
	\bibinfo{author}{Kaczorowski, D.} \& \bibinfo{author}{Wi{\'{s}}niewski, P.}
	\newblock \bibinfo{journal}{\bibinfo{title}{{Magnetoresistance in LuBi and YBi
				semimetals due to nearly perfect carrier compensation}}}.
	\newblock {\emph{\JournalTitle{Physical Review B}}}
	\textbf{\bibinfo{volume}{97}}, \bibinfo{pages}{235132}
	(\bibinfo{year}{2018}).
	
	\bibitem{Shekhar2015b}
	\bibinfo{author}{Shekhar, C.} \emph{et~al.}
	\newblock \bibinfo{journal}{\bibinfo{title}{{Extremely large magnetoresistance
				and ultrahigh mobility in the topological Weyl semimetal candidate NbP}}}.
	\newblock {\emph{\JournalTitle{Nat. Phys.}}} \textbf{\bibinfo{volume}{11}},
	\bibinfo{pages}{645} (\bibinfo{year}{2015}).
	
	\bibitem{Li2016f}
	\bibinfo{author}{Li, Y.} \emph{et~al.}
	\newblock \bibinfo{journal}{\bibinfo{title}{{Resistivity plateau and negative
				magnetoresistance in the topological semimetal TaSb$_2$}}}.
	\newblock {\emph{\JournalTitle{Phys. Rev. B}}} \textbf{\bibinfo{volume}{94}},
	\bibinfo{pages}{121115} (\bibinfo{year}{2016}).
	
	\bibitem{Singha2016a}
	\bibinfo{author}{Singha, R.}, \bibinfo{author}{Pariari, A.~K.},
	\bibinfo{author}{Satpati, B.} \& \bibinfo{author}{Mandal, P.}
	\newblock \bibinfo{journal}{\bibinfo{title}{{Large nonsaturating
				magnetoresistance and signature of nondegenerate Dirac nodes in ZrSiS}}}.
	\newblock {\emph{\JournalTitle{Proc. Natl. Acad. Sci.}}}
	\textbf{\bibinfo{volume}{114}}, \bibinfo{pages}{2468}
	(\bibinfo{year}{2017}).
	
	\bibitem{Hosen2017}
	\bibinfo{author}{Hosen, M.~M.} \emph{et~al.}
	\newblock \bibinfo{journal}{\bibinfo{title}{{Tunability of the topological
				nodal-line semimetal phase in ZrSi$X$-type materials ($X$=S, Se, Te)}}}.
	\newblock {\emph{\JournalTitle{Phys. Rev. B}}} \textbf{\bibinfo{volume}{95}},
	\bibinfo{pages}{161101} (\bibinfo{year}{2017}).
	
	\bibitem{Zhao2015}
	\bibinfo{author}{Zhao, Y.} \emph{et~al.}
	\newblock \bibinfo{journal}{\bibinfo{title}{{Anisotropic magnetotransport and
				exotic longitudinal linear magnetoresistance in WTe$_2$ crystals}}}.
	\newblock {\emph{\JournalTitle{Phys. Rev. B}}} \textbf{\bibinfo{volume}{92}},
	\bibinfo{pages}{041104} (\bibinfo{year}{2015}).
	
	\bibitem{Du2005}
	\bibinfo{author}{Du, X.}, \bibinfo{author}{Tsai, S.-W.},
	\bibinfo{author}{Maslov, D.~L.} \& \bibinfo{author}{Hebard, A.~F.}
	\newblock \bibinfo{journal}{\bibinfo{title}{{Metal-insulator-like behavior in
				semimetallic bismuth and graphite}}}.
	\newblock {\emph{\JournalTitle{Phys. Rev. Lett.}}}
	\textbf{\bibinfo{volume}{94}}, \bibinfo{pages}{166601}
	(\bibinfo{year}{2005}).
	
	\bibitem{Wang2015d}
	\bibinfo{author}{Wang, Y.~L.} \emph{et~al.}
	\newblock \bibinfo{journal}{\bibinfo{title}{{Origin of the turn-on temperature
				behavior in WTe$_2$}}}.
	\newblock {\emph{\JournalTitle{Phys. Rev. B}}} \textbf{\bibinfo{volume}{92}},
	\bibinfo{pages}{180402} (\bibinfo{year}{2015}).
	
	\bibitem{Zeng2016}
	\bibinfo{author}{Zeng, L.-K.} \emph{et~al.}
	\newblock \bibinfo{journal}{\bibinfo{title}{{Compensated semimetal LaSb with
				unsaturated magnetoresistance}}}.
	\newblock {\emph{\JournalTitle{Phys. Rev. Lett.}}}
	\textbf{\bibinfo{volume}{117}}, \bibinfo{pages}{127204}
	(\bibinfo{year}{2016}).
	
	\bibitem{Niu2016}
	\bibinfo{author}{Niu, X.~H.} \emph{et~al.}
	\newblock \bibinfo{journal}{\bibinfo{title}{{Presence of exotic electronic
				surface states in LaBi and LaSb}}}.
	\newblock {\emph{\JournalTitle{Phys. Rev. B}}} \textbf{\bibinfo{volume}{94}},
	\bibinfo{pages}{165163} (\bibinfo{year}{2016}).
	
	\bibitem{Pavlosiuk2016f}
	\bibinfo{author}{Pavlosiuk, O.}, \bibinfo{author}{Swatek, P.} \&
	\bibinfo{author}{Wi{\'{s}}niewski, P.}
	\newblock \bibinfo{journal}{\bibinfo{title}{{Giant magnetoresistance,
				three-dimensional Fermi surface and origin of resistivity plateau in YSb
				semimetal}}}.
	\newblock {\emph{\JournalTitle{Sci. Rep.}}} \textbf{\bibinfo{volume}{6}},
	\bibinfo{pages}{38691} (\bibinfo{year}{2016}).
	
	\bibitem{Kumar2016}
	\bibinfo{author}{Kumar, N.} \emph{et~al.}
	\newblock \bibinfo{journal}{\bibinfo{title}{{Observation of
				pseudo-two-dimensional electron transport in the rock salt-type topological
				semimetal LaBi}}}.
	\newblock {\emph{\JournalTitle{Phys. Rev. B}}} \textbf{\bibinfo{volume}{93}},
	\bibinfo{pages}{241106} (\bibinfo{year}{2016}).
	
	\bibitem{Ghimire2016}
	\bibinfo{author}{Ghimire, N.~J.}, \bibinfo{author}{Botana, A.~S.},
	\bibinfo{author}{Phelan, D.}, \bibinfo{author}{Zheng, H.} \&
	\bibinfo{author}{Mitchell, J.~F.}
	\newblock \bibinfo{journal}{\bibinfo{title}{{Magnetotransport of single
				crystalline YSb}}}.
	\newblock {\emph{\JournalTitle{J. Phys. Condens. Matter}}}
	\textbf{\bibinfo{volume}{28}}, \bibinfo{pages}{235601}
	(\bibinfo{year}{2016}).
	
	\bibitem{Han2017a}
	\bibinfo{author}{Han, F.} \emph{et~al.}
	\newblock \bibinfo{journal}{\bibinfo{title}{{Separation of electron and hole
				dynamics in the semimetal LaSb}}}.
	\newblock {\emph{\JournalTitle{Phys. Rev. B}}} \textbf{\bibinfo{volume}{96}},
	\bibinfo{pages}{125112} (\bibinfo{year}{2017}).
	
	\bibitem{Xu2017f}
	\bibinfo{author}{Xu, J.} \emph{et~al.}
	\newblock \bibinfo{journal}{\bibinfo{title}{{Origin of the extremely large
				magnetoresistance in the semimetal YSb}}}.
	\newblock {\emph{\JournalTitle{Phys. Rev. B}}} \textbf{\bibinfo{volume}{96}},
	\bibinfo{pages}{075159} (\bibinfo{year}{2017}).
	
	\bibitem{Chen2016a}
	\bibinfo{author}{Chen, F.~C.} \emph{et~al.}
	\newblock \bibinfo{journal}{\bibinfo{title}{{Extremely large magnetoresistance
				in the type-II Weyl semimetal MoTe$_2$}}}.
	\newblock {\emph{\JournalTitle{Phys. Rev. B}}} \textbf{\bibinfo{volume}{94}},
	\bibinfo{pages}{235154} (\bibinfo{year}{2016}).
	
	\bibitem{Kumar2017c}
	\bibinfo{author}{Kumar, N.} \emph{et~al.}
	\newblock \bibinfo{journal}{\bibinfo{title}{{Extremely high magnetoresistance
				and conductivity in the type-II Weyl semimetals WP$_2$ and MoP$_2$}}}.
	\newblock {\emph{\JournalTitle{Nat. Commun.}}} \textbf{\bibinfo{volume}{8}},
	\bibinfo{pages}{1642} (\bibinfo{year}{2017}).
	
	\bibitem{Wang2016l}
	\bibinfo{author}{Wang, X.} \emph{et~al.}
	\newblock \bibinfo{journal}{\bibinfo{title}{{Evidence of both surface and bulk
				Dirac bands and anisotropic nonsaturating magnetoresistance in ZrSiS}}}.
	\newblock {\emph{\JournalTitle{Advanced Electronic Materials}}}
	\textbf{\bibinfo{volume}{2}}, \bibinfo{pages}{1600228}
	(\bibinfo{year}{2016}).
	
	\bibitem{Ziman1972}
	\bibinfo{author}{Ziman, J.}
	\newblock \emph{\bibinfo{title}{Principles of the Theory of Solids}}
	(\bibinfo{publisher}{Cambridge University Press}, \bibinfo{year}{1972}).
	
	\bibitem{Shoenberg1984}
	\bibinfo{author}{Shoenberg, D.}
	\newblock \emph{\bibinfo{title}{Magnetic Oscillations in Metals}}
	(\bibinfo{publisher}{Cambridge University Press}, \bibinfo{year}{1984}).
	
	\bibitem{Ando2013}
	\bibinfo{author}{Ando, Y.}
	\newblock \bibinfo{journal}{\bibinfo{title}{{Topological insulator
				materials}}}.
	\newblock {\emph{\JournalTitle{J. Phys. Soc. Jpn.}}}
	\textbf{\bibinfo{volume}{82}}, \bibinfo{pages}{102001}
	(\bibinfo{year}{2013}).
	
	\bibitem{Wang2016k}
	\bibinfo{author}{Wang, C.~M.}, \bibinfo{author}{Lu, H.~Z.} \&
	\bibinfo{author}{Shen, S.~Q.}
	\newblock \bibinfo{journal}{\bibinfo{title}{{Anomalous phase shift of quantum
				oscillations in 3D topological semimetals}}}.
	\newblock {\emph{\JournalTitle{Phys. Rev. Lett.}}}
	\textbf{\bibinfo{volume}{117}}, \bibinfo{pages}{077201}
	(\bibinfo{year}{2016}).
	
	\bibitem{Li2018d}
	\bibinfo{author}{Li, C.} \emph{et~al.}
	\newblock \bibinfo{journal}{\bibinfo{title}{{Rules for phase shifts of quantum
				oscillations in topological nodal-line semimetals}}}.
	\newblock {\emph{\JournalTitle{Phys. Rev. Lett.}}}
	\textbf{\bibinfo{volume}{120}}, \bibinfo{pages}{146602}
	(\bibinfo{year}{2018}).
	
	\bibitem{Taskin2011b}
	\bibinfo{author}{Taskin, A.~A.} \& \bibinfo{author}{Ando, Y.}
	\newblock \bibinfo{journal}{\bibinfo{title}{{Berry phase of nonideal Dirac
				fermions in topological insulators}}}.
	\newblock {\emph{\JournalTitle{Phys. Rev. B}}} \textbf{\bibinfo{volume}{84}},
	\bibinfo{pages}{035301} (\bibinfo{year}{2011}).
	
	\bibitem{Hu2016}
	\bibinfo{author}{Hu, J.} \emph{et~al.}
	\newblock \bibinfo{journal}{\bibinfo{title}{{$\pi$ Berry phase and Zeeman
				splitting of Weyl semimetal TaP}}}.
	\newblock {\emph{\JournalTitle{Sci. Rep.}}} \textbf{\bibinfo{volume}{6}},
	\bibinfo{pages}{18674} (\bibinfo{year}{2016}).
	
	\bibitem{Gao2017}
	\bibinfo{author}{Gao, W.} \emph{et~al.}
	\newblock \bibinfo{journal}{\bibinfo{title}{{Extremely large magnetoresistance
				in a topological semimetal candidate pyrite PtBi$_2$}}}.
	\newblock {\emph{\JournalTitle{Phys. Rev. Lett.}}}
	\textbf{\bibinfo{volume}{118}}, \bibinfo{pages}{256601}
	(\bibinfo{year}{2017}).
	
	\bibitem{Wang2017b}
	\bibinfo{author}{Wang, A.} \emph{et~al.}
	\newblock \bibinfo{journal}{\bibinfo{title}{{Large magnetoresistance in the
				type-II Weyl semimetal WP$_2$}}}.
	\newblock {\emph{\JournalTitle{Phys. Rev. B}}} \textbf{\bibinfo{volume}{96}},
	\bibinfo{pages}{121107} (\bibinfo{year}{2017}).
	
	\bibitem{Liang2014}
	\bibinfo{author}{Liang, T.} \emph{et~al.}
	\newblock \bibinfo{journal}{\bibinfo{title}{{Ultrahigh mobility and giant
				magnetoresistance in the Dirac semimetal Cd$_3$As$_2$.}}}
	\newblock {\emph{\JournalTitle{Nat. Mater.}}} \textbf{\bibinfo{volume}{14}},
	\bibinfo{pages}{280--284} (\bibinfo{year}{2015}).
	
	\bibitem{Luo2015a}
	\bibinfo{author}{Luo, Y.} \emph{et~al.}
	\newblock \bibinfo{journal}{\bibinfo{title}{{Hall effect in the extremely large
				magnetoresistance semimetal WTe$_2$}}}.
	\newblock {\emph{\JournalTitle{Appl. Phys. Lett.}}}
	\textbf{\bibinfo{volume}{107}}, \bibinfo{pages}{182411}
	(\bibinfo{year}{2015}).
	
	\bibitem{Hu2017b}
	\bibinfo{author}{Hu, J.} \emph{et~al.}
	\newblock \bibinfo{journal}{\bibinfo{title}{{Nearly massless Dirac fermions and
				strong Zeeman splitting in the nodal-line semimetal ZrSiS probed by de
				Haas–van Alphen quantum oscillations}}}.
	\newblock {\emph{\JournalTitle{Phys. Rev. B}}} \textbf{\bibinfo{volume}{96}},
	\bibinfo{pages}{045127} (\bibinfo{year}{2017}).
	
	\bibitem{Pippard2009}
	\bibinfo{author}{Pippard, A.}
	\newblock \emph{\bibinfo{title}{Magnetoresistance in Metals}}
	(\bibinfo{publisher}{Cambridge University Press}, \bibinfo{year}{2009}).
	
	\bibitem{Miller1996}
	\bibinfo{author}{Miller, D.} \& \bibinfo{author}{Laikhtman, B.}
	\newblock \bibinfo{journal}{\bibinfo{title}{{Longitudinal magnetoresistance of
				superlattices caused by barrier inhomogeneity}}}.
	\newblock {\emph{\JournalTitle{Phys. Rev. B}}} \textbf{\bibinfo{volume}{54}},
	\bibinfo{pages}{10669} (\bibinfo{year}{1996}).
	
	\bibitem{Argyres1956}
	\bibinfo{author}{Argyres, P.~N.} \& \bibinfo{author}{Adams, E.~N.}
	\newblock \bibinfo{journal}{\bibinfo{title}{{Longitudinal magnetoresistance in
				the quantum limit}}}.
	\newblock {\emph{\JournalTitle{Phys. Rev.}}} \textbf{\bibinfo{volume}{104}},
	\bibinfo{pages}{900} (\bibinfo{year}{1956}).
	
	\bibitem{Stroud1976}
	\bibinfo{author}{Stroud, D.} \& \bibinfo{author}{Pan, F.~P.}
	\newblock \bibinfo{journal}{\bibinfo{title}{{Effect of isolated inhomogeneities
				on the galvanomagnetic properties of solids}}}.
	\newblock {\emph{\JournalTitle{Phys. Rev. B}}} \textbf{\bibinfo{volume}{13}},
	\bibinfo{pages}{1434} (\bibinfo{year}{1976}).
	
	\bibitem{Pal2010}
	\bibinfo{author}{Pal, H.~K.} \& \bibinfo{author}{Maslov, D.~L.}
	\newblock \bibinfo{journal}{\bibinfo{title}{{Necessary and sufficient condition
				for longitudinal magnetoresistance}}}.
	\newblock {\emph{\JournalTitle{Phys. Rev. B}}} \textbf{\bibinfo{volume}{81}},
	\bibinfo{pages}{214438} (\bibinfo{year}{2010}).
	
	\bibitem{Faruseth1965}
	\bibinfo{author}{Faruseth, S.}, \bibinfo{author}{Selte, K.} \&
	\bibinfo{author}{Kjekshus, A.}
	\newblock \bibinfo{journal}{\bibinfo{title}{{Redetermined crystal structures of
				NiTe$_2$, PdTe$_2$, PtS$_2$, PtSe$_2$, and PtTe$_2$}}}.
	\newblock {\emph{\JournalTitle{Acta Chem. Scand.}}}
	\textbf{\bibinfo{volume}{19}}, \bibinfo{pages}{257} (\bibinfo{year}{1965}).
	
\end{thebibliography}
\end{document}